\documentclass[aps,prl,groupedaddress]{revtex4}
\usepackage{mitpress}
\usepackage{amsmath}
\usepackage{graphicx}%
\usepackage{amsfonts}%
\usepackage{amssymb}

\newdimen\dummy
\dummy=\oddsidemargin
\begin{document}

\title{Quantum Computing and Dynamical Quantum Models}
\author{Scott Aaronson}
\affiliation{University of California, Berkeley\\aaronson@cs.berkeley.edu}

\begin{abstract}
A \textit{dynamical quantum model} assigns an eigenstate to a specified observable
even when no measurement is made, and gives a stochastic evolution rule for
that eigenstate. \ Such a model yields a distribution over classical histories of a
quantum state. \ We study what can be computed by sampling from that
distribution, i.e., by examining an observer's entire history. \ We show that, relative to an oracle, one can solve problems in
polynomial time that are intractable even for quantum computers; and\ can
search an $N$-element list in order $N^{1/3}$\ steps (though not fewer).
\pacs{03.65.Ta, 03.67.Lx}
\end{abstract}
\maketitle


Given a system with known Hamiltonian and initial state, nonrelativistic
quantum theory specifies the probability of measuring an observable in a given
eigenstate at each time $t\geq0$. \ It does not, however, yield multiple-time or
transition probabilities \cite{bd,barrett,dc,dieks}: that is, what is the
probability that an observable assumes value $\left|  \psi_{2}\right\rangle
$\ at $t_{2}$, given that it assumed $\left|  \psi_{1}\right\rangle $\ at an
earlier time $t_{1}$ (though was not measured at $t_{1}$)?

This question, we argue, is not easily ignored in scenarios wherein a
Hamiltonian $H$\ is coherently applied to a system that contains an observer.
\ Granting the possibility of macroscopic coherence, one might try to avoid
the question as follows. \ In a typical experiment, one keeps records of
observations. \ If the records were stored where $H$ could not affect them,
then they would inhibit the interference that $H$ would otherwise produce.
\ If, on the other hand, records were not kept, or were themselves subject to
$H$, then at time $t_{2}$\ one would lose the right to ask which eigenstate
was observed\ at $t_{1}$, since at $t_{2}$\ this event has no meaning outside
of records available at $t_{2}$. \ The difficulty with this account is that it
is unclear what a `prediction' could mean without some notion of multiple-time
probabilities independent of records. \ Indeed, the `outcome of an experiment,' the `output of a
computation,' and the `utility of a decision' all seem to presuppose an
observer or collection of observers persisting over time. \ (See
\cite[p.126]{barrett}\ and \cite[p.135-6]{bell}\ for related criticisms.)

The question of multiple-time probabilities arises in any interpretation of
quantum theory that treats observers as physical systems that can be placed in
superposition. \ This includes many-worlds interpretations,\ modal
interpretations, and the Bohm interpretation, though not
`explicit-collapse'\ interpretations. \ The Bohm interpretation asserts an
answer to our question, but applies only to a particular setting: it assumes
not only the form of the guiding equation, but also a\ state space (the
positions and momenta of particles in Euclidean space) and a preferred
observable (position). \ We take a more abstract perspective, which
allows arbitrary finite state spaces, and does not commit to any one
observable or dynamics.

In this Letter we initiate the study of multiple-time probabilities from the standpoint
of quantum computing. \ Our main result is that an observer, if given
access to past eigenstates, could solve problems efficiently that are
believed to be intractable even for quantum computers. \  Although such access is not permitted by
quantum theory, an observer might wish to {\it calculate} a probable history of
eigenstates, given a dynamics, initial state, and Hamiltonian. \ Under weak assumptions, we show that this task is infeasible
even for moderately-sized systems, meaning those for which a quantum computer could efficiently sample eigenstates at {\it single} times.

We define a \textit{dynamical quantum model} to be a function which, given a
pure or mixed state $\rho$ in $N$ dimensions, a unitary $U$ acting on $\rho$,
and a von Neumann observable $V$ with $N$ possible outcomes, specifies (for
all $i,j$) a probability $p_{ij}$\ that $V$ assumes value $i$ before $U$ is
applied to $\rho$ (time $t_{1}$), and $j$ after $U$ is applied (time $t_{2}$).
\ The $p_{ij}$'s must marginalize to the single-time probabilities implied by
quantum theory: that is, the diagonal entries of $V\rho V^{-1}$\ at $t_{1}$,
and of $VU\rho U^{-1}V^{-1}$\ at $t_{2}$. \ It is immediate that there exists
a dynamical model (the simplest one, which we call the \textit{product
dynamics} or $\mathcal{PD}$, takes the distribution over $V$ at $t_{2}$ to be
independent of that at $t_{1}$), and that there are infinitely many
nonequivalent models.

Let a quantum computer have initial state $\left|  0\right\rangle ^{\otimes
n}$, and suppose we apply a sequence $\mathcal{U}=\left(  U_{1},\ldots
,U_{T}\right)  $\ of unitary operations, each of which can be implemented in
\textsf{BQP}, or bounded-error quantum polynomial time. \ (See \cite{nc}\ for
background on \textsf{BQP} and other computational complexity classes.) \ Let $V$ be the observable corresponding to the standard
(computational) basis. \ We consider a \textit{history} $H=\left(
v_{1},\ldots,v_{T}\right)  $ of $V$, which chooses a particular eigenstate of
$V$ at each time step: namely, $v_{k}$\ immediately after $U_{k}$\ is applied
to the state $U_{k-1}\cdots U_{1}\left|  0\right\rangle ^{\otimes n}$. \ Then
any dynamics $\mathcal{D}$ yields a distribution $\Omega\left(  \mathcal{U}%
,\mathcal{D}\right)  $\ over histories. \ Observe that $\Omega$\ is a Markov
distribution; that is, each $v_{k}$\ is independent of the other $v_{l}$'s
conditioned on $v_{k-1}$\ and $v_{k+1}$. \ Sampling a history from $\Omega$ is
at least as difficult as simulating a polynomial-time quantum
computation,\ for sampling from the marginal distribution over any $v_{k}$\ is
equivalent to simulating a standard-basis measurement of $U_{k}\cdots
U_{1}\left|  0\right\rangle ^{\otimes n}$. \ But could sampling a history
enable one to solve problems that are intractable even for quantum computers?

To separate this question from any particular dynamics, we introduce a
complexity class $\mathsf{DQP}$, or dynamical quantum polynomial-time.
\ Informally, $\mathsf{DQP}$\ consists of those problems solvable in
polynomial time by sampling histories under \textit{any} dynamical model, so
long as it satisfies locality and symmetry conditions to be discussed. \ We then show that
$\mathsf{SZK}\subseteq\mathsf{DQP}$, where $\mathsf{SZK}$, statistical zero
knowledge, is a classical complexity class containing several problems that
have resisted efficient quantum algorithms---including graph isomorphism,
nonabelian hidden subgroup, and approximate shortest lattice vector. \ Already
this suggests that $\mathsf{BQP}\neq\mathsf{DQP}$, i.e., that our dynamical
quantum computing model is strictly more powerful than the usual quantum computing
model. \ However, we give stronger evidence, of the kind typically sought in
computer science. \ We recently obtained \cite{aaronson}\ a lower bound of
order $n^{1/5}$\ on the number of oracle queries needed by a quantum computer
to solve the `collision problem,' that of deciding whether a function
$f:\left\{  1,\ldots,n\right\}  \rightarrow\mathbb{Z}$ is one-to-one or
two-to-one.\ \ (Shi \cite{shi} has improved this bound to order $n^{1/3}$,
which is optimal.) \ But the collision problem, which abstractly models
$\mathsf{SZK}$, can be solved in a \textit{constant} number of queries using a
dynamical model. \ Formalizing this intuition, we show in \cite{aaronson} that
there exists an oracle $A$ for which $\mathsf{SZK}^{A}\nsubseteq
\mathsf{BQP}^{A}$, and therefore $\mathsf{BQP}^{A}\neq\mathsf{DQP}^{A}$.

As is usual in quantum computing, we assume a Hilbert space $\mathcal{H}_{N}$
of finite dimension $N$,\ and discretize time into steps of equal length
$\tau$. \ Other authors \cite{bd}\ have considered dynamics in a
continuous-time setting. \ It might be thought that our restriction to
discrete time introduces a drawback, that the dynamics depend not just on the
initial state and Hamiltonian but also on the choice of $\tau$. \ For example,
two Hadamard gates applied in succession seem to correspond to two random
transitions when considered separately, but to a permutation (namely the
identity permutation) when considered jointly. \ However, an analogous problem
occurs in the continuous-time setting. \ There, letting $t\rightarrow0$ be the
length of a time interval being considered, there is still a free parameter
$d\tau/dt$ on which the dynamics depend.

For simplicity, we consider the dynamics of only a single time-independent
observable $V$, and assume those dynamics at each time $t$ to depend only on
the state and Hamiltonian at $t$ (it is easy to show that they cannot depend
on the Hamiltonian only).\ \ By the Kochen-Specker theorem we cannot assign
values noncontextually to every observable, let alone specify their transition
probabilities. \ We could consider a subset $S$ of observables that contains
no Kochen-Specker contradiction, but even then we could not apply a dynamical
model independently to each observable in $S$ without in general violating
noncontextuality. \ The case of time-dependent observables was considered in
\cite{bd} and elsewhere.

Formally,\ a dynamical quantum model\ is fully characterized by a family of
functions, $\left\{  \mathcal{D}_{N}\right\}  _{N\geq1}:\mathcal{H}_{N}\times
U\left(  N\right)  ^{2}\rightarrow S\left(  N\right)  $, which map a pure or
mixed state $\rho\in\mathcal{H}_{N}$, a unitary $U\in U\left(  N\right)  $,
and an orthonormal basis $V=v_{1},\ldots,v_{N}\in U\left(  N\right)  $\ onto a
singly stochastic matrix $S\in S\left(  N\right)  $. \ We sometimes suppress
the dependence on $N$. \ Let $\left(  M\right)  _{ij}$\ denote the entry in
the $i^{th}$\ column and $j^{th}$\ row of $M$. \ Then $\left(  S\right)
_{ij}$ is the probability that the observable corresponding to $V$ takes value
$v_{j}$ after $U$\ is applied to $\rho$, conditioned on $V$ taking value
$v_{i}$\ before $U$ is applied.\ \ Any dynamics must satisfy the conditions of
\textit{unitary invariance}---for all unitary changes of basis $W$%
,\vspace{0pt}%
\[
\vspace{0pt}\mathcal{D}\left(  \rho,U,V\right)  =\mathcal{D}\left(  W\rho
W^{-1},WUW^{-1},WVW^{-1}\right)  ,
\]
and \textit{marginalization}---for all $j\in\left\{  1,\ldots,N\right\}
$,\vspace{0pt}%
\[%
{\textstyle\sum\nolimits_{i}}
\left(  S\right)  _{ij}\left(  \rho\right)  _{ii}=\left(  U\rho U^{-1}\right)
_{jj}\text{.}%
\]
Because of invariance, we will henceforth take $V=I$\ and consider
$\mathcal{D}$\ as a function of $\rho$\ and $U$\ only.

Three additional conditions we desire are \textit{symmetry},
\textit{robustness}, and \textit{locality}. \ We say that $\mathcal{D}$ is
symmetric if it is invariant under relabeling of basis states: more precisely,
for all permutation matrices $P$ and $Q$,\vspace{0pt}%
\[
\vspace{0pt}\mathcal{D}\left(  P\rho P^{-1},QUP^{-1}\right)  =Q\mathcal{D}%
\left(  \rho,U\right)  P^{-1}\text{.}%
\]
Also, $\mathcal{D}$\ is robust if it is insensitive to sufficiently small
errors (which, in particular, implies continuity): for all polynomials $p$,
there exists a polynomial $q$\ such that for all $N$, $\rho\in\mathcal{H}_{N}%
$, and $U\in U\left(  N\right)  $,\vspace{0pt}%
\[
\vspace{0pt}\left\|  \mathcal{D}_{N}\left(  \rho,U\right)  -\mathcal{D}%
_{N}\left(  \rho^{\ast},U^{\ast}\right)  \right\|  \leq1/p\left(  N\right)
\]
where $\left\|  M\right\|  =\max_{ij}\left|  \left(  M\right)  _{ij}\right|
$, whenever $\left\|  \rho-\rho^{\ast}\right\|  \leq1/q\left(  N\right)  $ and
$\left\|  U-U^{\ast}\right\|  \leq1/q\left(  N\right)  $. \ Robustness will
not be needed for our results, but is often demanded of a computational model.

In the interest of generality, we did not assume $\mathcal{H}_{N}$\ to have a
particular tensor product structure. \ Thus, we define locality by
partitioning the basis states\ into `blocks,' between which $U$ can never
produce interference. \ Call $L\subseteq\left\{  1,\ldots,N\right\}  $ a
\textit{block} if $\left(  U\right)  _{ik}=0$ for all $i\in L$\ and $k\notin
L$, and a \textit{minimal block} if no $L^{\ast}\subset L$ is a block. \ Note
that the minimal blocks are disjoint. \ Then $\mathcal{D}$\ is local if it
acts separately on each minimal block: more formally,%
\[
\left(  S\right)  _{ij}=\mathcal{D}_{\left|  L\right|  }\left(  \rho_{L}%
,U_{L}\right)
\]
for all minimal blocks $L$ and $i,j\in L$, where $U_{L}$\ is the $L\times
L$\ submatrix of $U$, and $\rho_{L}$\ is the $L\times L$\ submatrix of $\rho
$\ normalized to have trace $1$. \ We do not claim that the locality condition
implies relativistic causality. \ For example, if $\rho_{AB}$\ is a bipartite
state and $U_{A}$ and $U_{B}$\ act only on $A$ and $B$ respectively, then
locality does not imply \textit{commutativity} in the sense that\vspace{0pt}%
\begin{align*}
&  \mathcal{D}\left(  U_{A}\rho_{AB}U_{A}^{-1},U_{B}\right)  \mathcal{D}%
\left(  \rho_{AB},U_{A}\right) \\
&  \vspace{0pt}=\mathcal{D}\left(  U_{B}\rho_{AB}U_{B}^{-1},U_{A}\right)
\mathcal{D}\left(  \rho_{AB},U_{B}\right)  \text{.}%
\end{align*}
We raise as an open question whether there exists a dynamical model satisfying
robustness, locality, and commutativity. \ See \cite{dc}\ for a more detailed
analysis of causality in dynamical models.

The product dynamics $\mathcal{PD}$\ is unsatisfactory because it does not
satisfy locality. \ Dieks \cite{dieks}\ proposed partitioning the basis
vectors into minimal blocks and applying $\mathcal{PD}$\ separately to each.
\ The resulting \textit{Dieks dynamics}, $\mathcal{DD}$,\ satisfies locality
and commutativity, but not robustness, since the minimal blocks are sensitive
to arbitrarily small changes to $U$.

We introduce a dynamical model, the \textit{Schr\"{o}dinger dynamics} or
$\mathcal{SD}$, that satisfies robustness and locality. \ Commutativity is
satisfied for unentangled states but not for entangled ones. \ Constructing
$\mathcal{SD}$\ involves solving a system of nonlinear equations, which were
first studied in the continuous case by Schr\"{o}dinger\ \cite{schrodinger}.
\ The existence and uniqueness of a solution was shown under broad conditions
by Nagasawa \cite{nagasawa}. \ In the discrete case, where the problem is
known as $\left(  r,c\right)  $\textit{-scaling}, efficient algorithms are
known for finding the solution (\cite{lsw} and references therein).

The idea is repeatedly to tweak $U$ to bring it closer to\ a stochastic matrix
that satisfies the marginalization condition. \ The first step is to replace
each entry of $U$ by its squared magnitude, obtaining $U^{\left(  0\right)  }%
$\ such that $\left(  U^{\left(  0\right)  }\right)  _{ij}=\left|  \left(
U\right)  _{ij}\right|  ^{2}$. \ We wish to make the $i^{th}$ column of the
matrix sum to $\left(  \rho\right)  _{ii}$, and the $j^{th}$ row sum to
$\left(  U\rho U^{-1}\right)  _{jj}$ for all $i,j\in\left\{  1,\ldots
,N\right\}  $. \ The stochastic matrix $S$ mapping $\operatorname*{diag}%
\left(  \rho\right)  $ to $\operatorname*{diag}\left(  U\rho U^{-1}\right)
$\ is then readily obtained by normalizing each column to sum to $1$. \ Here
`normalizing' means multiplying by a scalar.

The algorithm is iterative. \ For each $t\geq0$\ we obtain\ $U^{\left(
2t+1\right)  }$\ by normalizing each column $i$ of $U^{\left(  2t\right)  }%
$\ to sum to $\left(  \rho\right)  _{ii}$; likewise we obtain $U^{\left(
2t+2\right)  }$\ by normalizing each row $j$ of $U^{\left(  2t+1\right)  }%
$\ to sum to $\left(  U\rho U^{-1}\right)  _{jj}$. \ We claim that (1) the
limit $U^{\left(  \infty\right)  }$\ of this iteration exists, and (2) the
resulting diagonal matrices $A$ and $B$ such that $U^{\left(  \infty\right)
}=AU^{\left(  0\right)  }B$\ are unique up to scalar multiples. \ It is known
\cite{lsw} that both claims are implied by the following `flow condition':
there exists a nonnegative matrix $M$\ such that $\sum_{k}\left(  M\right)
_{ik}=\left(  \rho\right)  _{ii}$\ for all $i$, $\sum_{k}\left(  M\right)
_{kj}=\left(  U\rho U^{-1}\right)  _{jj}$\ for all $j$, and $\left(  M\right)
_{ij}=0$\ whenever $\left(  U\right)  _{ij}=0$. \ Surprisingly, the flow
condition always holds if $U$ is unitary (we omit the proof).

Clearly $\mathcal{SD}$\ satisfies locality, since $U_{ij}^{\left(
\infty\right)  }=0$\ whenever $U_{ij}=0$.\ \ It is shown in \cite{lsw}\ that
for any polynomial $p$, the iterative algorithm converges to within
$1/p\left(  N\right)  $\ precision in polynomial time. \ Using this one can
show that $\mathcal{SD}$ satisfies robustness also.

We now define the complexity class $\mathsf{DQP}$. \ For a dynamics
$\mathcal{D}$, let $\mathcal{O}\left(  \mathcal{D}\right)  $\ be an oracle
that takes as input a sequence $\mathcal{U}=\left(  U_{1},\ldots,U_{T}\right)
$\ of quantum circuits, and returns as output a sample $\left(  v_{1}%
,\ldots,v_{T}\right)  $\ from the history distribution $\Omega\left(
\mathcal{U},\mathcal{D}\right)  $\ as defined previously. \ Then let
$\mathsf{DQP}\left(  \mathcal{D}\right)  =\mathsf{BQP}^{\mathcal{O}\left(
\mathcal{D}\right)  }$ (i.e. $\mathsf{BQP}$\ with oracle access to
$\mathcal{O}\left(  \mathcal{D}\right)  $), and let $\mathsf{DQP}$\ be the set
of languages that are in $\mathsf{DQP}\left(  \mathcal{D}\right)  $\ for all
$\mathcal{D}$\ satisfying symmetry and locality. \ Other reasonable classes
could be defined---for example, we could allow only classical queries to
$\mathcal{O}\left(  \mathcal{D}\right)  $, or only one query instead of
multiple ones---but such distinctions are a subject for complexity theory
rather than physics. \ The best upper bound we know of is $\mathsf{DQP}%
\subseteq\mathsf{P}^{\mathsf{\#P}}$, from the Dieks dynamics.

Let us see why $\mathsf{SZK}\subseteq\mathsf{DQP}$. \ Sahai and Vadhan
\cite{sv}\ showed that, to simulate $\mathsf{SZK}$, it suffices to solve the
following \textit{statistical difference} ($SD$) problem. \ Suppose
deterministic classical polynomial-time algorithm $P_{i}$\ (for $i\in\left\{
0,1\right\}  $) returns output $Y_{i}\left(  X\right)  \in\left\{
0,1\right\}  ^{n+1}$ distributed according to $\Lambda_{i}=\left(
p_{Y,i}\right)  $,\ when given an input $X$ chosen uniformly from $\left\{
0,1\right\}  ^{n}$. \ Then decide whether $\Lambda_{0}$\ and $\Lambda_{1}%
$\ are `$\varepsilon$-close'\ or `$\varepsilon$-far'---that is, whether%
\[
\vspace{0pt}\left\|  \Lambda_{0}-\Lambda_{1}\right\|  =%
{\textstyle\sum\nolimits_{Y}}
\left|  p_{Y,0}-p_{Y,1}\right|  /2
\]
is less than $\varepsilon$\ or greater than $1-\varepsilon$ for some
$\varepsilon>0$, given that one of these is the case. \ As an example, let
$G_{0}$\ and $G_{1}$\ be graphs, and let $\Lambda_{i}$\ be the uniform
distribution over all permutations of $G_{i}$. \ Then $\Lambda_{0}$ and
$\Lambda_{1}$\ are $0$-close (that is, identical) if $G_{0}$\ and $G_{1}$\ are
isomorphic, and are $0$-far (disjoint) otherwise.\ \ It follows that testing
isomorphism of graphs is reducible to $SD$, and hence is in $\mathsf{SZK}$.

In the special case where $P_{0}$\ and $P_{1}$ are one-to-one, the
$\mathsf{DQP}$\ algorithm consists simply of three quantum circuits,
$U_{1},U_{2},$ and $U_{3}$. \ First $U_{1}$\ transforms $\left|
0\right\rangle ^{\otimes n}$\ to $\left(  \left|  \Phi_{0}\right\rangle
+\left|  \Phi_{1}\right\rangle \right)  /\sqrt{2}$, where \vspace{0pt}%
\[
\vspace{0pt}\left|  \Phi_{i}\right\rangle =2^{-n/2}%
{\textstyle\sum\nolimits_{X\in\left\{  0,1\right\}  ^{n}}}
\left|  i\right\rangle \left|  X\right\rangle \left|  Y_{i}\left(  X\right)
\right\rangle
\]
for a control bit $\left|  i\right\rangle $ (henceforth $\left|  Y_{i}\left(
X\right)  \right\rangle $\ is abbreviated $\left|  Y\right\rangle $). \ Then
$U_{2}$\ applies a bitwise Fourier transform to $\left|  i\right\rangle
\left|  X\right\rangle $ (that is, a Hadamard gate on each bit), and $U_{3}%
$\ does the same, returning the state to $U_{1}\left|  0\right\rangle
^{\otimes n}$. \ Intuitively, this is analogous to measuring $\left|
Y\right\rangle $, and then making multiple `non-collapsing' measurements of
$\left|  i\right\rangle $ to see whether it contains one value or a
superposition of two values. \ In the former case we conclude that
$\Lambda_{0}\ $and $\Lambda_{1}$\ are $\varepsilon$-far; in the latter that
they are $\varepsilon$-close. \ The technical part is to show that this
algorithm works under \textit{any} symmetric local model.

Let $v_{k}=\left|  i_{k}\right\rangle \left|  X_{k}\right\rangle \left|
Y\right\rangle $\ be the value of $V$ immediately after $U_{k}$\ is applied.
\ First suppose $\Lambda_{0}\ $and $\Lambda_{1}$\ are $\varepsilon$-far.
\ Then because $P_{0}$\ and $P_{1}$\ are one-to-one, $v_{1}$'s `counterpart'
$\left|  \urcorner i_{1}\right\rangle \left|  X_{1}^{\left(  \urcorner\right)
}\right\rangle \left|  Y\right\rangle $\ has zero amplitude in $U_{1}\left|
0\right\rangle ^{\otimes n}$\ with probability at least $1-\varepsilon$, where
`$\urcorner$'\ denotes negation. \ In that case, the state of $\left|  i\right\rangle $
conditioned on $\left|  Y\right\rangle $\ is $\left|  i_{1}\right\rangle $.
\ Since $U_{2}$ and $U_{3}$\ do not act on $\left|  Y\right\rangle $ and
$U_{3}U_{2}$ is the identity, it follows by locality that $i_{1}=i_{3}$.

Second, suppose $\Lambda_{0}\ $and $\Lambda_{1}$\ are $\varepsilon$-close.
\ Define binary vectors $a=i_{1}\circ X_{1}$, $b=\urcorner i_{1}\circ
X_{1}^{\left(  \urcorner\right)  }$, and $c=i_{2}\circ X_{2}$ in
$\mathbb{Z}_{2}^{n+1}$, where `$\circ$'\ denotes concatenation. \ Then
$\left|  a\right\rangle \left|  Y\right\rangle $\ and\ $\left|  b\right\rangle
\left|  Y\right\rangle $ have equal amplitude with probability at least
$1-\varepsilon$. \ Recall that the Fourier transform $F$\ maps $\left|
a\right\rangle $\ onto $2^{-n/2}%
{\textstyle\sum\nolimits_{c}}
\left(  -1\right)  ^{a\cdot c}\left|  c\right\rangle $\ and similarly for
$\left|  b\right\rangle $. \ Thus, the only $\left|  c\right\rangle $\ that
have nonzero amplitude in $U_{2}U_{1}\left|  0\right\rangle ^{\otimes n}$ are
those for which $a\cdot c\equiv b\cdot c\left(  \operatorname{mod}2\right)  $.
$\ $We wish to show that $F$ is symmetric under some permutation of
eigenstates that interchanges $a$ with $b$ while leaving $c$ fixed. \ Suppose
we had an invertible matrix $M$ over $\mathbb{Z}_{2}^{n+1}$\ such that $Ma=b$,
$Mb=a$, and $M^{T}c=c$. \ Then define two permutations $\sigma,\tau$\ over
binary vectors by $\sigma\left(  a\right)  =Ma$\ and $\tau\left(  c\right)
=\left(  M^{T}\right)  ^{-1}c$, so that%
\[
\sigma\left(  a\right)  \cdot\tau\left(  c\right)  \equiv a\cdot c\left(
\operatorname{mod}2\right)
\]
for all $a,c$. \ Since the $\left(  a,c\right)  $\ entry of $F$ is
$2^{-\left(  n+1\right)  /2}\left(  -1\right)  ^{a\cdot c}$, this implies that
$F$ is symmetric under application of $\sigma$\ to its input eigenstates and
$\tau^{-1}$\ to its output eigenstates. \ We argue that such an $M$ exists so
long as $a$ and $b$ are nonzero (which they almost certainly are). \ For let
$w$ and $z$ be unit vectors, and let $L$ be an invertible matrix over
$\mathbb{Z}_{2}^{n+1}$ such that $Lw=a$\ and $Lz=b$. \ Let $Q$ be the
permutation matrix that interchanges $w$ and $z$ while leaving all other unit
vectors fixed. \ Then set $M=LQL^{-1}$. \ Clearly $Ma=b$ and $Mb=a$. \ Also,
$a\cdot c=b\cdot c$ implies $w^{T}L^{T}c=z^{T}L^{T}c$, so the $w$ and $z$
entries of $L^{T}c$\ are equal, and thus $Q^{T}\left(  L^{T}c\right)  =L^{T}%
c$, implying $M^{T}c=c$.

By the symmetry condition, it follows that $\left(  S\right)  _{ca}=\left(
S\right)  _{cb}=1/2$, where $\left(  S\right)  _{ca}$\ is the probability that
$v_{3}=\left|  a\right\rangle \left|  Y\right\rangle $\ and $\left(  S\right)
_{cb}$\ that $v_{3}=\left|  b\right\rangle \left|  Y\right\rangle $, both
conditioned on $v_{2}=\left|  c\right\rangle \left|  Y\right\rangle $. \ Thus,
there is a $1/2$ probability that $i_{1}\neq i_{3}$.

For general $P_{0}$\ and $P_{1}$, we can reduce to the one-to-one case by
appending a register $\left|  h\left(  i\circ X\right)  \right\rangle
$\ to\ $\left|  \Phi_{i}\right\rangle $, on which $U_{2}$\ and $U_{3}$\ do not
act. \ Here $h$ is chosen uniformly at random among all `hash functions'
mapping $\mathbb{Z}_{2}^{n+1}$\ to $\left\{  1,\ldots,K\right\}  $, for some
range size $K$. \ Let $n_{0}=\left|  P_{0}^{-1}\left(  Y\right)  \right|
$\ be the number of $X$\ such that $P_{0}\left(  X\right)  =Y$,\ and similarly
define $n_{1}$.\ \ Then assuming that $\Lambda_{0}\ $and $\Lambda_{1}$\ are
$\varepsilon$-close, $\left|  n_{0}/n_{1}-1\right|  <4\varepsilon$\ with probability at least $3/4$ over the choice of $Y$, by Markov's inequality. \ After applying $U_{1}$, we apply $U_{2}%
$\ and $U_{3}$\ in succession $n$ times, initially with $K=1$ and each time
thereafter setting $K$ to twice its previous value and recomputing $h$.
\ Define $a=i_{1}\circ X_{1}$\ as before. \ Then we want there to exist a
unique counterpart $b=\urcorner i_{1}\circ X_{1}^{\left(  \urcorner\right)  }%
$\ such that $h\left(  a\right)  =h\left(  b\right)  $, but no $a^{\ast}%
=i_{1}\circ X_{1}^{\ast}$\ such that
$h\left(  a\right)  =h\left(  a^{\ast}\right)  $.\ \ Letting $\alpha=n_{1}/K$,
this joint event (call it $E$) occurs with probability%
\[
\left(  1-1/K\right)  ^{n_{0}+n_{1}}n_{1}/K\approx\alpha e^{-2\alpha\left(
1\pm2\varepsilon\right)  }%
\]
over the choice of $h$. \ This is bounded away from $0$ when $\alpha\in\left[
1,2\right]  $. \ When $E$ does occur, the analysis for the one-to-one case
applies, and establishes that $v_{1}$\ and its counterpart are both observed
with $1/2$ probability.

The algorithm for searching an unordered list of $N$ items in order $N^{1/3}%
$\ queries is conceptually similar. \ Assume for simplicity that there is a
unique marked state $\left|  j\right\rangle $ that we are trying to find.
\ The first step is to apply $N^{1/3}$\ iterations of Grover's search
algorithm \cite{grover}, thereby boosting the probability of observing
$\left|  j\right\rangle $ to order $N^{-1/3}$. \ The next step is to `juggle'
the observable $V$ as uniformly as possible, so that after order $N^{1/3}%
$\ steps, with high probability $V$ has visited $\left|  j\right\rangle $\ at
least once. \ Then $j$ can be found by inspecting the classical history
$\left(  v_{1},\ldots,v_{T}\right)  $. \ Again the technical part is to show
that this can be done in any symmetric local model, and again the primary
tools are a hash function (to reduce the problem of juggling $V$ among many
eigenstates to that of juggling it between two), and the bitwise Fourier
transform (to juggle). \ Details are omitted due to space limitations.

The $N^{1/3}$\ bound is easily seen to be optimal under \textit{any} dynamical
model. \ Bennett et al. \cite{bbbv}\ showed that, if $\Psi^{\left(  t\right)
}\left(  X\right)  $\ is an algorithm's state after $t$ queries to an $N$-item
list $X$, then by changing one item of $X$ we can obtain a list $X^{\ast}$
such that $\left\|  \Psi^{\left(  t\right)  }\left(  X\right)  -\Psi^{\left(
t\right)  }\left(  X^{\ast}\right)  \right\|  \precsim t^{2}/N$\ in trace
distance. \ It follows by the union bound that, if $T\ll N^{1/3}$\ queries are
made, then the probability that the $X\rightarrow X^{\ast}$ change affects the
history $\left(  v_{1},\ldots,v_{T}\right)  $\ is of order $\sum_{t=1}%
^{T}t^{2}/N\ll1$. \ Hence, there exists an oracle $A$ relative to which
\textsf{NP}-complete problems are not efficiently solvable in dynamical
models; that is, \textsf{NP}$^{A}\nsubseteq\mathsf{DQP}^{A}\left(
\mathcal{D}\right)  $ for any $\mathcal{D}$. \ This result supports the
intuition that dynamical models are somehow more `physically reasonable' than
(for example) nonlinear quantum models, which {\it would} enable \textsf{NP}%
-complete and even \textsf{\#P}-complete problems to be solved in polynomial
time \cite{al}. \ Although our model grants an observer access to her entire
history within a quantum system, it does not allow her to record histories in
superposition, or otherwise to influence the system in a way contrary to
quantum theory.

I thank Ronald de Wolf, Umesh Vazirani, John Preskill, Guido Bacciagaluppi,
Avi Wigderson, and Dennis Dieks for helpful discussions. \ Supported by an NSF
Graduate Fellowship and by DARPA grant F30602-01-2-0524.

\end{document}